\RequirePackage[modulo]{lineno}
\documentclass[twocolumn,showpacs,aps,floatfix,prd,nofootinbib]{revtex4-1}
\usepackage{graphicx}
\usepackage{amsmath}
\usepackage{amsmath}
\usepackage{lineno}

\newcommand{\BABARPubNumber}{18/006}
\newcommand{\SLACPubNumber}{17318}

\input babarsym

\def\systcr {12\%}
 
\usepackage{color}
\usepackage{hyperref}
\usepackage[hang]{footmisc}
\usepackage{scrextend}
 
\begin{document}
\begin{minipage}{.45\linewidth}
\begin{flushleft}
\babar\--PUB-\BABARPubNumber \\
SLAC-PUB-\SLACPubNumber \\
\end{flushleft}
\end{minipage}
\title{\large \bf
\boldmath
Measurement of the $\gamma^{\star}\gamma^{\star} \to \eta'$ transition form factor 
}
%
\author{J.~P.~Lees}
\author{V.~Poireau}
\author{V.~Tisserand}
\affiliation{Laboratoire d'Annecy-le-Vieux de Physique des Particules (LAPP), Universit\'e de Savoie, CNRS/IN2P3,  F-74941 Annecy-Le-Vieux, France}
\author{E.~Grauges}
\affiliation{Universitat de Barcelona, Facultat de Fisica, Departament ECM, E-08028 Barcelona, Spain }
\author{A.~Palano}
\affiliation{INFN Sezione di Bari and Dipartimento di Fisica, Universit\`a di Bari, I-70126 Bari, Italy }
\author{G.~Eigen}
\affiliation{University of Bergen, Institute of Physics, N-5007 Bergen, Norway }
\author{D.~N.~Brown}
\author{Yu.~G.~Kolomensky}
\affiliation{Lawrence Berkeley National Laboratory and University of California, Berkeley, California 94720, USA }
\author{M.~Fritsch}
\author{H.~Koch}
\author{T.~Schroeder}
\affiliation{Ruhr Universit\"at Bochum, Institut f\"ur Experimentalphysik 1, D-44780 Bochum, Germany }
% ...
\collaboration{The \babar\ Collaboration}
\noaffiliation

\begin{abstract}
We study the process $e^+e^-\to e^+e^- \eta'$ in the double-tag mode and
measure for the first time the $\gamma^{\star}\gamma^{\star} \to \eta'$
transition form factor $F_{\eta'}(Q_1^2, Q_2^2)$ in the momentum-transfer
range \mbox{2 $<Q_1^2, Q_2^2<$ 60 GeV$^2$}. The analysis is based on
a data sample corresponding to an integrated luminosity of around  469 fb$^{-1}$ collected at the PEP-II $e^+e^-$ collider with the \babar\ detector at center-of-mass energies
near 10.6 GeV.
\end{abstract}
\maketitle

\section{\boldmath Introduction \label{intro}}
\begin{figure}[h]
\includegraphics[width=.4\textwidth]{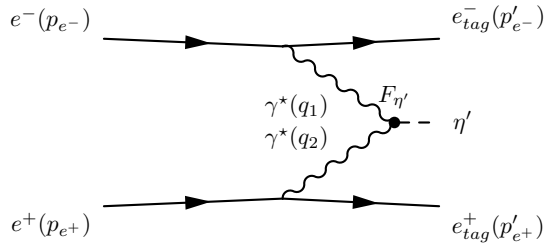} 
\caption{The diagram for the $e^+e^-\to e^+e^- \eta'$ process.
\label{diagram}}
\end{figure}
In this article, we report on the measurement of the 
$\gamma^{\star}\gamma^{\star} \to \eta'$ transition form factor (TFF) by 
using the two-photon-fusion reaction 
\begin{linenomath}
\begin{equation}
\label{eeeeetap}
e^+e^-\to e^+e^- \eta' \nonumber
\end{equation}
\end{linenomath}
illustrated by the diagram in Fig.~\ref{diagram}. 
The TFF is defined via the amplitude for the  
$\gamma^{\star}\gamma^{\star} \to \eta'$ transition
\begin{linenomath} 
\begin{equation}\label{FF}
T = - i 4\pi \alpha \epsilon_{\mu \nu \beta \gamma} \varepsilon_{1}^{\mu} \varepsilon_{2}^{\nu} q_1^{\beta} q_2^{\gamma} F_{\eta'}(Q_1^2, Q_2^2),
\end{equation}
\end{linenomath}
where $\alpha$ is the  fine structure constant,
$\epsilon_{\mu \nu \alpha \beta}$ is the totally antisymmetric Levi-Civita 
tensor, $\varepsilon_{1,2}$ and $q_{1,2}$ are the polarization
vectors and four-momenta, respectively, of the space-like photons, $Q_{1,2}^2=-q_{1,2}^2$,
and $F_{\eta'}(Q_1^2, Q_2^2)$ is the  transition form factor.

We measure the differential cross section of the process $e^+e^-\to e^+e^- \eta'$
in the double-tag mode, in which both scattered electrons\footnote{
Unless otherwise specified, we use the term ``electron'' for either an electron or a positron.} 
are detected (tagged). The tagged electrons emit highly
off-shell photons with momentum transfers
$q_{e^{+}}^2 = -Q_{e^{+}}^2 = (p_{e^{+}}-p'_{e^{+}})^2$ and 
$q_{e^{-}}^2 = - Q_{e^{-}}^2 = (p_{e^{-}}-p'_{e^{-}})^2$,
where $p_{e^{\pm}}$ and $p^\prime_{e^{\pm}}$ are the four-momenta, respectively, of the initial- and final-state
electrons. We measure for the first time 
 $F_{\eta'}(Q_1^2, Q_2^2)$ in the kinematic region with two highly 
off-shell photons $2 < Q_1^2, Q_2^2 < 60$ GeV$^2$.
The $\eta'$ transition form factor $F_{\eta'}(Q^2, 0)$
in the space-like momentum transfer region and 
in the single-tag mode  was measured 
 in several previous experiments \cite{bPLUTO,bTPC,bCELLO,bCLEO,old_etastudy}. The most precise data at large $Q^2$ were obtained by the CLEO~\cite{bCLEO} experiment, and then by the \babar~\cite{old_etastudy} experiment, in the momentum transfer ranges $1.5 < Q^2 < 30$ GeV$^2$ and $4 < Q^2 < 40$ GeV$^2$, respectively.
  
\begin{figure*}
\begin{center}
\includegraphics[width=.4\textwidth]{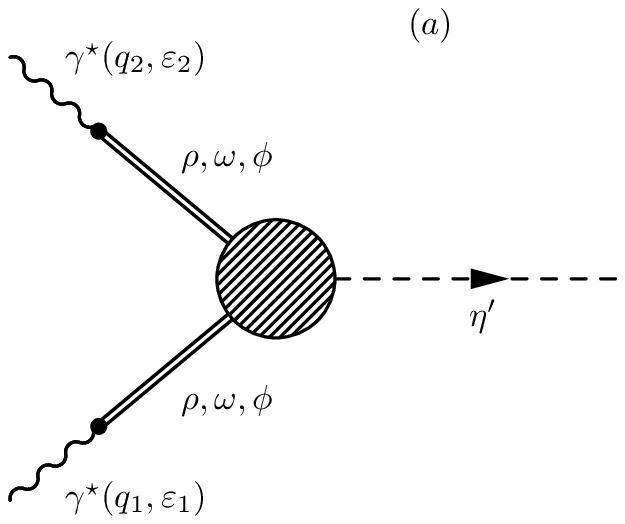}
\includegraphics[width=.4\textwidth]{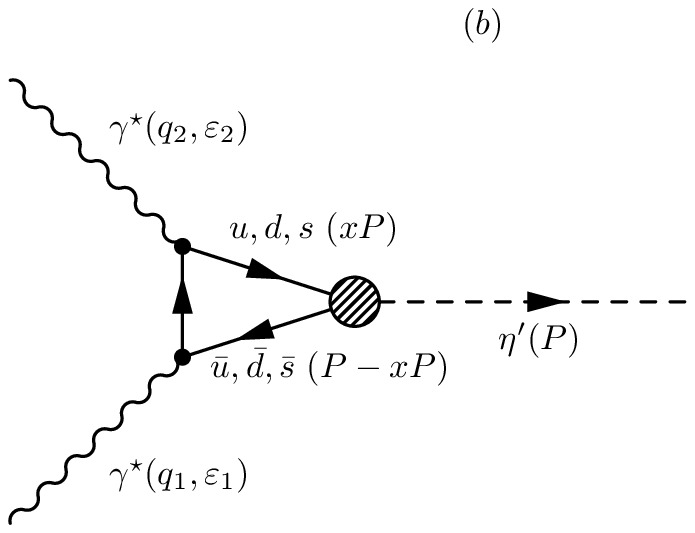}
\caption{The Feynman diagrams for the process $\gamma^{\star}\gamma^{\star}\to \eta'$ in the VDM (a) and pQCD (b).
\label{VMD_QCD}}
\end{center}
\end{figure*}

Many theoretical models exist for the description of the TFFs of
pseudoscalar mesons, $F_{P}(Q_1^2, 0)$ and $F_{P}(Q_1^2, Q_2^2)$ (see for example Refs.~\cite{bsumrules,QCD_inspired_model,bKroll,bVMD1}). 
Measurement of the TFF at large $Q_1^2$ and $Q_2^2$ allows the predictions of models inspired by perturbative QCD (pQCD) to be distinguished from those of the vector dominance model (VDM)~\cite{bVMD2,bVMD3,bVMD}. The tree-level diagrams for VDM and pQCD approaches are shown in Fig.~\ref{VMD_QCD}. In the case of only one off-shell photon, both classes of models predict the same asymptotic dependence $F_{P}(Q^2, 0)\sim 1/Q^2$ as $Q^2 \to \infty$, while for two off-shell photons the asymptotic predictions are quite different, $F(Q_1^2, Q_2^2)\sim 1/(Q_1^2+Q_2^2)$ for pQCD, and $F(Q_1^2, Q_2^2)\sim 1/(Q_1^2Q_2^2)$ for the VDM model.

\section{\boldmath Theoretical approach to the form factor 
$F_{\eta'}(Q_1^2, Q_2^2)$.\label{Theoretical_approach}}

As a consequence of $\eta-\eta'$ mixing, the $\eta'$ wave function can be 
represented as the superposition of two quark-flavor 
states~\cite{et_etap_mixing}:
\begin{linenomath}
\begin{equation}
|\eta'\rangle = \sin{\phi}|n\rangle + \cos{\phi}|s\rangle,
\label{l6}
\end{equation}
\end{linenomath}
where
\begin{linenomath}
\begin{equation}
|n\rangle=\frac{1}{\sqrt{2}}\left(|\bar{u}u\rangle + |\bar{d}d\rangle\right),\:
|s\rangle = |\bar{s}s\rangle.
\label{l6a}
\end{equation}
\end{linenomath}
For the mixing angle $ \phi$ we use the value $\phi = (37.7 \pm 0.7)^\circ$~\cite{et_etap_mixing1}. 
The $\eta'$ transition form factor is related to the form factors for the
$|n\rangle$ and $|s\rangle$ states through
\begin{linenomath}
\begin{equation}\label{l5} 
F_{\eta'} = \sin{\phi} F_{n} + \cos{\phi} F_{s}.  
\end{equation}
\end{linenomath}
For large values of momentum transfer, pQCD predicts that the form factors $F_{n}$ and 
$F_{s}$ can be represented as a convolution of a hard scattering amplitude
$T_H$ and a non-perturbative meson distribution amplitude (DA) 
$\phi_{n,s}$:
\begin{linenomath}
\begin{equation}\label{l4} 
F_{n,s}(Q_{1}^2,Q_{2}^2) = \int^1_0 T_{H}(x,Q_{1}^2,Q_{2}^2,\mu) \phi_{n,s}(x,\mu) dx,
\end{equation}
\end{linenomath}
where $x$ is the longitudinal momentum fraction of the quark struck by the virtual photon in the hard scattering process. For the renormalization scale $\mu$, we take $\mu^2 = Q^2=Q_{1}^2 + Q_{2}^2$
as proposed in Ref.~\cite{BRAATEN} and for its asymptotic form $\phi_{n,s}$~\cite{Brodsky_Lepage}
\begin{linenomath}
\begin{equation}
\phi_{n,s} = 2C_{n,s}f_{n,s}6x(1-x)\left(1 + O(\Lambda_{QCD}^2/\mu^2)\right),
\end{equation}
\end{linenomath}
where the charge factors are $C_{n} = 5/(9\sqrt{2})$ and $C_{s} = {1}/{9}$,
the weak decay constants for the $|n\rangle$ and $|s\rangle$ states are
$f_n = (1.08 \pm 0.04)f_{\pi}$ and $f_s = (1.25 \pm 0.08)f_{\pi}$~\cite{et_etap_mixing1},
 $f_{\pi} = 130.4 \pm 0.2~\rm MeV$ is the pion decay constant, and
$\Lambda_{QCD}$ is the QCD scale parameter. 

In the case of two highly off-shell photons, $T_{H}(x,Q_{1}^2,Q_{2}^2)$ can
be represented as
\begin{eqnarray}
\label{l7}
T_{H}(x,Q_{1}^2,Q_{2}^2) = \frac{1}{2}
\frac{1}{xQ_{1}^{2} + (1-x)Q_{2}^{2}}
\\ \cdot \Bigg(1 + C_F \frac{\alpha_s(\mu^2)}{2\pi} t(x,Q_{1}^2,Q_{2}^2) \Bigg)  +
(x \to 1-x) \\ 
+ O(\alpha_s^2) + O(\Lambda_{QCD}^4/Q^4),
\end{eqnarray}
where ($x \to 1 - x$) stands for the first term
with replacement of $x$ by $1 - x$, $\alpha_s(\mu^2)$ is the QCD coupling strength, and $C_F = (n_c^2-1)/(2n_c) = {4}/{3}$ is a color factor. The expression for the next-to-leading order (NLO) component 
$t(x,Q_{1}^2,Q_{2}^2)$ can be found in Ref.~\cite{BRAATEN}, while the leading-order expression corresponds to $t(x,Q_{1}^2,Q_{2}^2) = 0$.
Combining Eqs.~(\ref{l5} -- \ref{l7}) we obtain the pQCD prediction for 
$F_{\eta'}(Q_{1}^2,Q_{2}^2)$ at large $Q_1^2$ and $Q_2^2$:
\begin{widetext}
\begin{equation}\label{adsadasdasd}
F_{\eta'}(Q_{1}^2,Q_{2}^2) = \left(\frac{5\sqrt{2}}{9} f_n \sin{\phi} + 
\frac{2}{9} f_s \cos{\phi} \right)
\int^1_0 dx   \frac{1}{2}\frac{6x(1-x)}{xQ_{1}^{2} + (1-x)Q_{2}^{2}} 
\left(1 + C_F \frac{\alpha_s(\mu^2)}{2\pi} t(x,Q_{1}^2,Q_{2}^2)\right) +
(x \to 1-x),  
\end{equation}
\end{widetext}

Significant effort has been invested to determine the DAs of pseudoscalar mesons at
intermediate values of momentum transfer~\cite{BRAATEN,Brodsky_Lepage,TFF_in_Q2_limit,Brodsky2011,Chernyak}. 
In contrast to the case of one off-shell photon, the TFF for two off-shell photons
is almost insensitive to the shape of the DA,
because the amplitude Eq.~(\ref{l7}) is finite at the endpoints $x=0$ and $x=1$.

According to the VDM model the TFF for the case of two off-shell photons is
\begin{linenomath}
\begin{equation}\label{lfff8}
F_{\eta'}(Q_{1}^2,Q_{2}^2) = \frac{F_{\eta'}(0,0)}{(1 + Q_1^2/\Lambda_P^2)(1 + Q_2^2/\Lambda_P^2)},
\end{equation}
\end{linenomath}
where $\Lambda_P$ is the pole mass parameter (see for example Ref.~\cite{bVMD}). In the case of the $\eta'$ meson, $\Lambda_P$ is found to be $849\pm6~\rm MeV/c^2$ from the approximation of $F_{\eta'}(Q^2, 0)$ with one off-shell photon~\cite{et_etap_mixing1}.
The value of $F_{\eta'}(0,0)$
can be obtained from the measured value of the $\eta'$ two-photon width
$\Gamma_{\eta^{\prime}\to 2\gamma} = 4.30 \pm 0.16$~keV~\cite{PDG2014}
using the formula~\cite{Brodsky_Lepage}:
\begin{linenomath}
\begin{equation}\label{l22}
F(0,0) =  \sqrt{\frac{4\Gamma_{\eta^{\prime}\to 2\gamma}}
{\pi\alpha^2 m_{\eta'}^3}} = 0.342 \pm 0.006~ \rm GeV^{-1}.
\end{equation}
\end{linenomath}
\section{\boldmath The \babar\ detector and data set \label{babar}}
The data used in this analysis were collected with the \babar\ detector at 
the \pep2\ asymmetric-energy \epem\ collider, at the SLAC National Accelerator Laboratory.
A total integrated luminosity of 468.6~\invfb~\cite{lumi} is used,
including 424.7~\invfb collected at the peak of $\Upsilon(4S)$ 
resonance and 43.9~\invfb collected 40 MeV below the resonance.

The \babar\ detector is described in detail elsewhere~\cite{Detector,Detector1}.
Charged particles are reconstructed using a tracking system, which includes
a silicon vertex tracker (SVT) and a drift chamber (DCH) inside a 1.5~T axial magnetic field. 
Separation of pions and kaons is accomplished by means of the 
detector of internally reflected Cherenkov light  and energy loss measurements
in the SVT and DCH. Photons are detected in the electromagnetic calorimeter
(EMC). Muon identification is provided by the instrumented flux return.

Signal $e^+e^-\to e^+e^- \eta^{\prime}$ events are simulated with the 
Monte Carlo (MC) event generator GGResRc~\cite{rGGResRc}. Because the 
$Q^{2}_{e^-}, Q^{2}_{e^+}$ distributions are peaked near zero, MC events
are generated with the requirement $Q^{2}_{e^-} (Q^{2}_{e^+}) > 2$ GeV$^2$. This
restriction corresponds to the limit of detector acceptance for the tagged
electrons.
The transition form factor in simulation is assumed to be constant.
The GGResRc event generator includes next-to-leading-order radiative 
corrections to the Born cross section calculated according to 
Ref.~\cite{abrakadabra}. In particular, it generates extra soft photons
emitted by the initial- and final-state electrons. 
The maximum center-of-mass (c.m.) energy of the photon emitted from the
initial state is required to be less than 0.05$\sqrt{s}$, where $\sqrt{s}$ 
is the $e^+e^-$ c.m.\ energy.
\section{\boldmath Event selection \label{eventselection}}
\begin{figure*}
\begin{minipage}[t]{0.48\textwidth}
\includegraphics[width=1\textwidth]{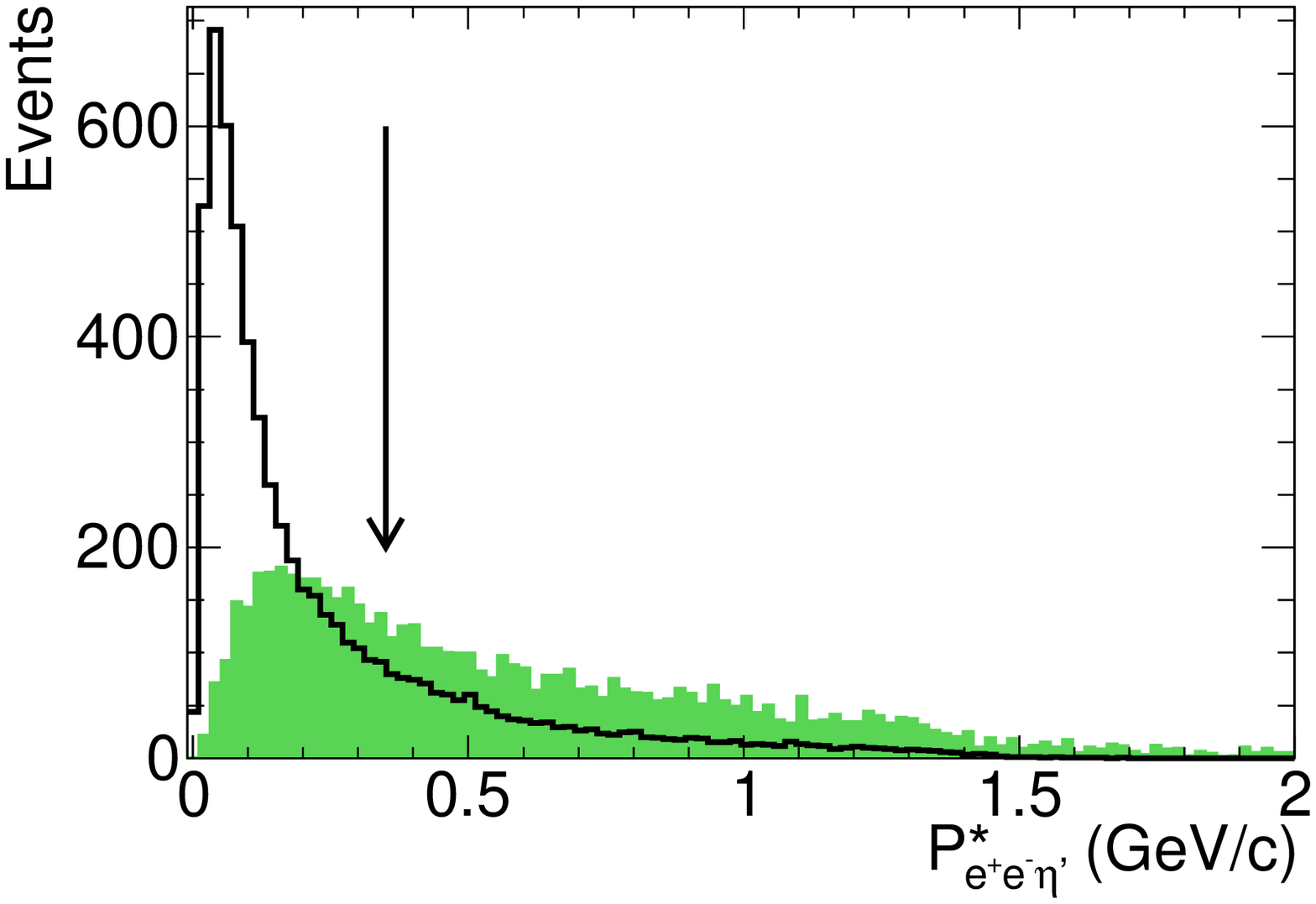}
\caption{Distribution of the total momentum of the $e^+e^-\eta'$ system 
in the c.m.\ frame.  The filled histogram shows the data. The open histogram represents MC simulation normalized to the number of events in data. Events with $P^{\star}_{e^+e^-\eta'}<0.35$ GeV/$c$ (indicated by the arrow) are retained for further analysis.
\label{deltaP}}
\end{minipage}
\hfill
\begin{minipage}[t]{0.48\textwidth}
\includegraphics[width=1\textwidth]{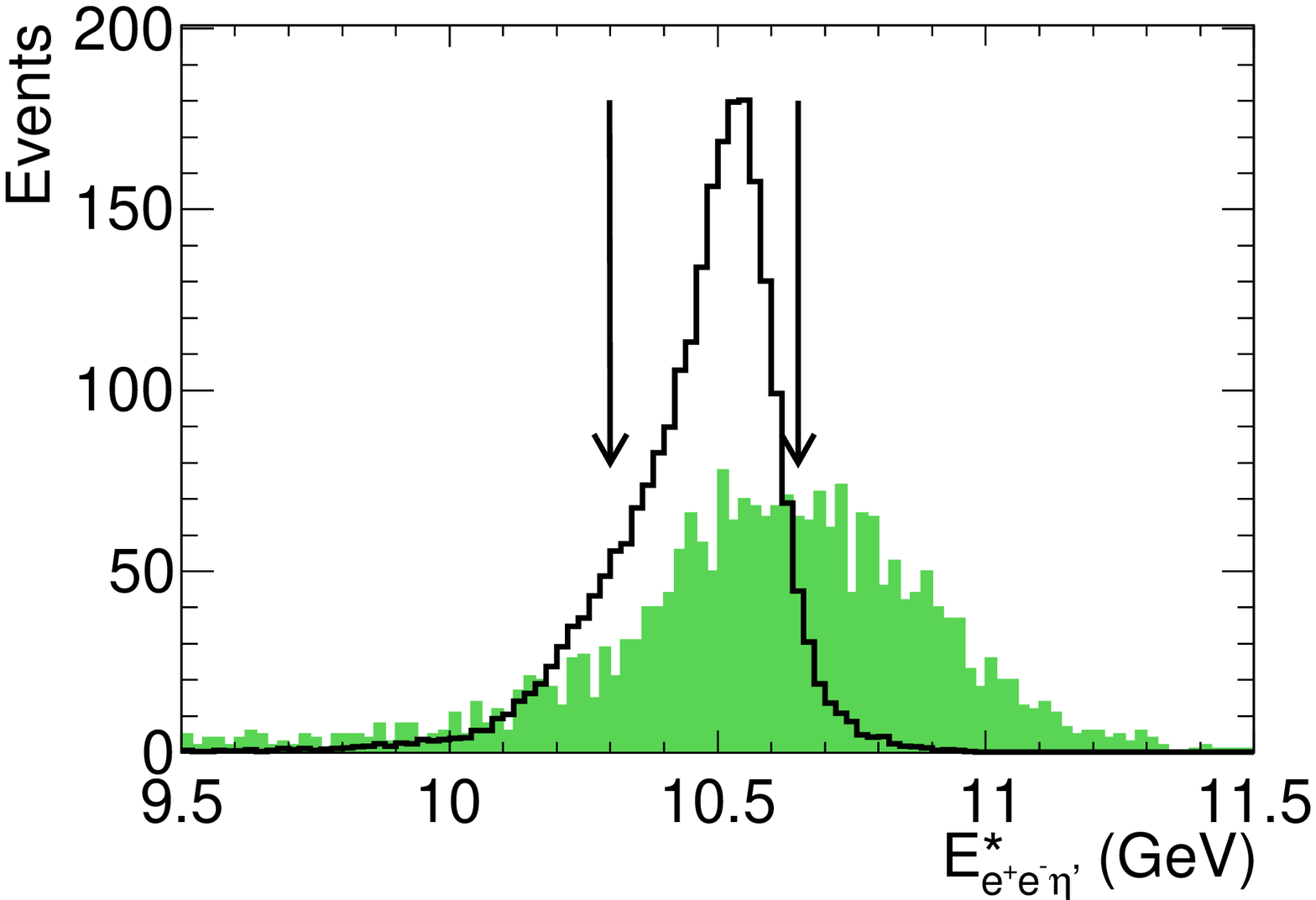}
\caption{Distribution of the total energy of the $e^+e^-\eta'$ system
in the c.m.\ frame. The filled histogram shows the data. The open histogram represents MC simulation normalized to the number of events in data. The arrows indicate the boundaries
  of the region used to select event candidates. 
\label{deltaE}}
\end{minipage}
\end{figure*}
\begin{figure*}
\begin{center}
\includegraphics[width=.6\textwidth]{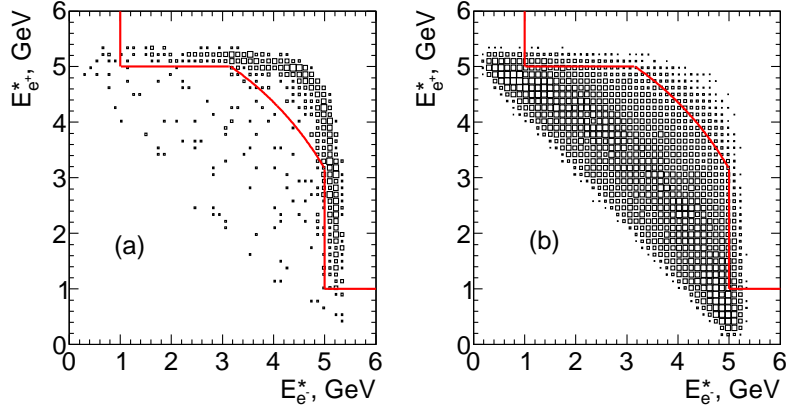} 
\caption{Distribution of the positron c.m.\ energy versus the electron c.m.\ energy for data (a) and simulated signal events (b). The lines indicate the boundary of the selection area. Events that lie above and to the right of the lines are rejected.
\label{Eel_Ep}}
\end{center}
\end{figure*}

The decay chain $\eta^{\prime} \to \pip \pim \eta \to \pip \pim 2\gamma$
is used to reconstruct the $\eta^\prime$ meson candidate. 

An initial sample of events with at least four tracks and two photon candidates is
selected. Tracks must have a point of closest approach to the nominal 
interaction point that is within 2.5 cm along the beam axis and less than
1.5 cm in the transverse plane. The track transverse momenta must be
greater than 50 MeV/$c$. Electrons  and pions are separated 
using a particle identification (PID) algorithm based on
information from the Cherenkov detector, EMC, and the tracking system. An event is required to contain two electron and two pion candidates. The electron PID efficiency is better than 98\%, with the 
pion misidentification probability below 10\%. 
The pion  PID efficiency is 98\%, with an electron misidentification
probability of about 7\%. 

To recover electron energy loss due to 
bremsstrahlung, the energy of all the calorimeter showers close to the electron
direction (within 35 and 50 mrad for the polar and azimuthal angle, 
respectively) is combined with the measured energy of the electron track.
The resulting c.m.\ energy of the electron candidate must be greater than 
0.2 GeV. 

The photon candidates are required to have an energy in the laboratory frame greater 
than 30 MeV. Two photon candidates are combined to form an $\eta$ candidate.
Their invariant mass is required to be in the 0.45--0.65 GeV$/c^2$ range.
We apply a kinematic fit to the two photons, with an $\eta$ mass constraint to improve 
the precision of their momentum measurement. An $\eta'$ candidate is formed
from a pair of oppositely charged pion candidates and an $\eta$ candidate.
The $\eta'$ candidate invariant mass must be in the range of 0.90--1.02 GeV$/c^2$. 

The final selection uses tagged electrons and is based on variables in the c.m.\ frame of the initial $e^+$ and $e^-$.
The total momentum of the reconstructed $e^+e^-\eta'$ system
($P^{\star}_{e^+e^-\eta'}$\footnote{The superscript asterisk indicates a quantity calculated in the $e^+e^-$ c.m.\ frame})
~must be less than 0.35 GeV/$c$. The distribution of the total momentum is shown in 
Fig.~\ref{deltaP} for data and simulated signal events.
The total energy of the $e^+e^-\eta'$ system must be in 
the range of 10.30--10.65 GeV as indicated by the arrows in Fig.~\ref{deltaE}.
To reject background from QED events, requirements on the energies of the 
detected electron and positron are applied. The two-dimensional distributions of
the electron c.m.\ energy versus the positron c.m.\ energy are shown in 
Fig.~\ref{Eel_Ep} for data and simulated signal events. The lines indicate the boundary of the selection area. Events that lie above and to the right of the lines are rejected.
\begin{figure*}
\begin{center}
\includegraphics[width=.8\textwidth]{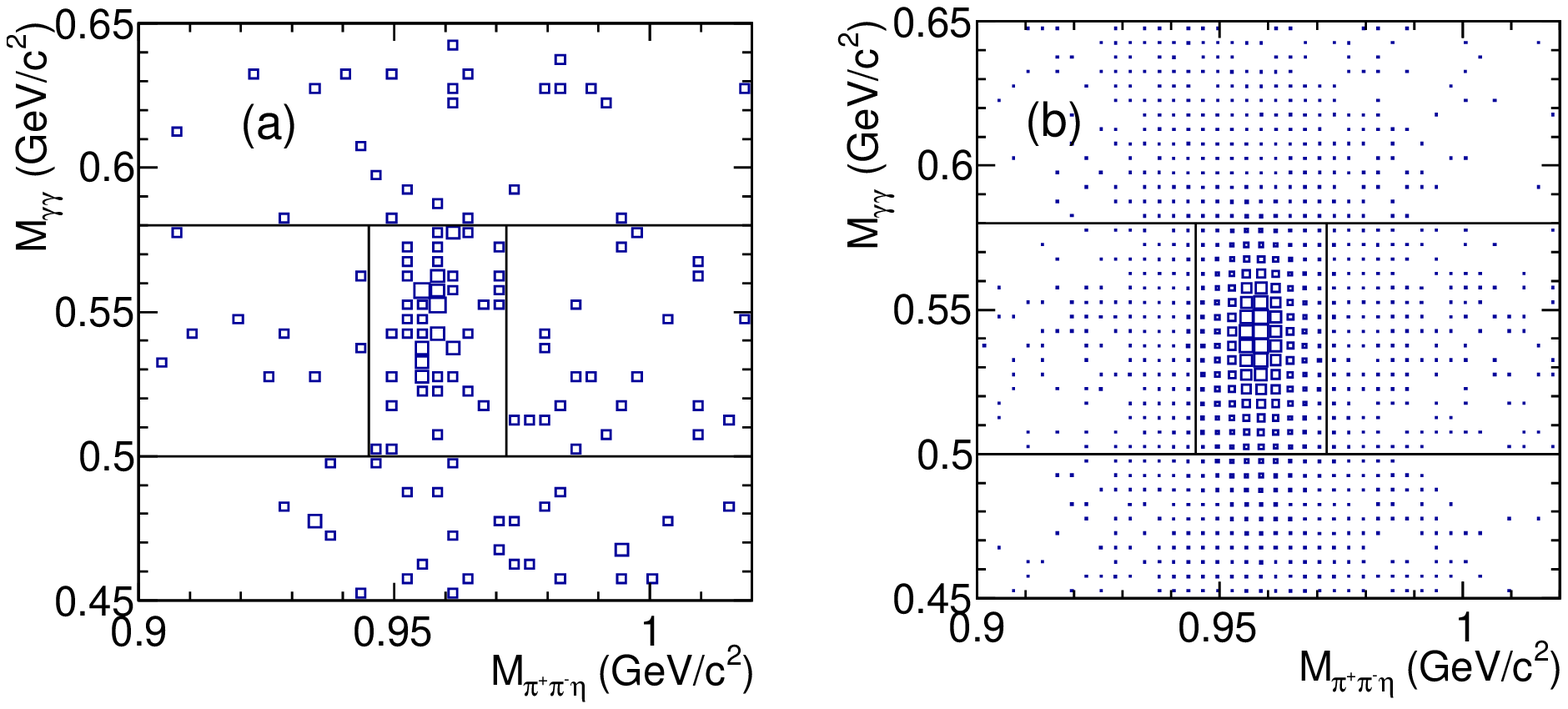}
\caption{Distribution of the $\eta$ candidate mass ($M_{\gamma\gamma}$)
versus the $\eta'$ candidate mass ($M_{\pi^+\pi^-\eta}$) for data (a) and 
signal MC simulation (b). The horizontal lines indicate the boundaries of
the selection condition applied. The vertical lines correspond to the restriction $0.945 < M_{\pi^+\pi^-\eta} < 0.972$  GeV/$c^2$ that is used for the plot of $Q_{e^-}^2$ versus $Q_{e^+}^2$ distribution in Fig.~\ref{Qm_Qp}.
\label{final}}
\end{center}
\end{figure*}
\begin{figure*}
\includegraphics[width=.45\textwidth]{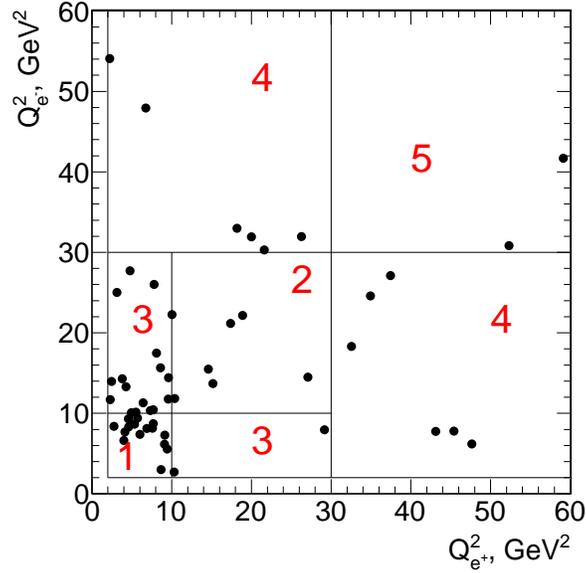} 
\caption{The $Q_{e^-}^2$ versus $Q_{e^+}^2$ distribution for  data events. The lines and numbers indicate the five regions used for the study of the dynamics of TFF a function of $Q_{e^-}^2$ and $Q_{e^+}^2$.
\label{Qm_Qp}}
\end{figure*}
 \begin{figure*}
 \begin{center}
 \includegraphics[width=0.7\textwidth]{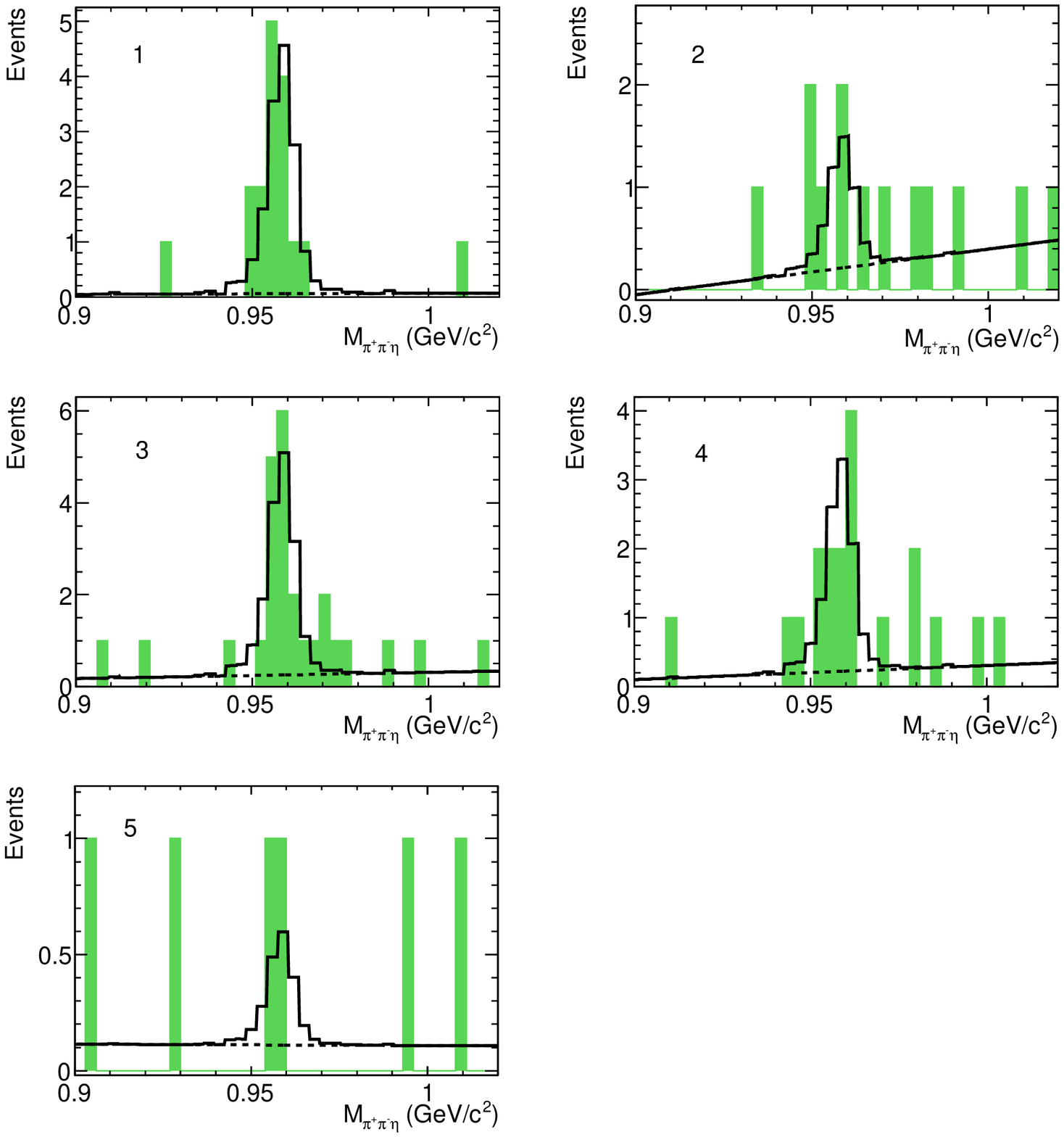}
 \caption{The $\pi^+\pi^-\eta$ mass spectra for data events from the five ($Q_1^2, Q_2^2$) regions of Fig.~\ref{Qm_Qp}. The open histograms are the fit results.
 The dashed lines represent fitted background.
 \label{Meta1}}
 \end{center}
 \end{figure*} 

The distribution of the $\eta$ candidate
mass versus the $\eta'$ one for the selected data and simulated signal
samples is shown in Fig.~\ref{final}. A clustering of events in the central region of the data distribution corresponds to the two-photon $\eta'$ production. To further suppress background we require that the invariant mass of the $\eta$ candidate be in the range 0.50--0.58 GeV/$c^2$, as shown by the horizontal lines in Fig.~\ref{final}. For events with more than one $\eta'$ or $e^{\pm}$ candidate (about 10\% of the selected events), the candidate with smallest absolute value of the total momentum of the $e^+e^-\eta'$ system in the c.m.\ frame is selected. 


Data events that pass all selection
criteria are divided into five ($Q_{e^-}^2$, $Q_{e^+}^2$) regions, as illustrated
on Fig.~\ref{Qm_Qp} for events with $0.945 < M_{\pi^+\pi^-\eta} < 0.972 $ GeV/c$^2$.
Because of the symmetry of the process under the exchange of the $e^-$ with the $e^+$, regions 3 and 4 each include two disjunct regions, mirror symmetric with respect to the diagonal. The number of signal events ($N_{\rm events}$)
in each ($Q_{e^-}^2$, $Q_{e^+}^2$) region is obtained from a fit to the
$\pi^+\pi^-\eta$ invariant mass spectrum with a sum of signal and background 
distributions as shown in Fig.~\ref{Meta1}. The signal line shape is obtained
from the signal simulation, while the background is assumed to be linear.
The fitted numbers of events for the five ($Q_{e^-}^2$, $Q_{e^+}^2$) regions
are listed in Table~\ref{cross_table0}. 
The total number of signal events is 
$46.2^{+8.3}_{-7.0}$. 
For the regions 2 and 5 we also use conservative estimates of the number of signal events as upper limits at 90\% C.L. using the
Feldman-Cousins approach~\cite{Feldman}.

To estimate the uncertainty related to the description
of the background, we repeat the fits using a quadratic background shape.
The deviation in the fitted number of signal events is 1.7\%. 
The uncertainty associated with the signal shape (3.3\%) is estimated by including into the
signal probability function a mass shift $\Delta M_{\pi^+\pi^-\eta} = -0.48$~MeV/c$^2$ and additional Gaussian smearing width $\sigma (M_{\pi^+\pi^-\eta}) = 1$~MeV/$c^2$. 
These parameters are obtained from our previous study
of $\gamma\gamma^* \to \eta'$ events~\cite{old_etastudy}, based on single-tagged events,
where the statistical precision was significantly larger.
The total systematic uncertainty (3.7\%) is obtained by adding
the individual terms in quadrature.

Following the methods developed in the single-tag analysis of Ref.~\cite{old_etastudy},
we have studied possible sources of peaking background:
$e^+e^-$ annihilation into hadrons, the two-photon process
$e^+e^- \to e^+e^-\eta' \pi^0$,  and the vector meson bremsstrahlung processes
$e^+e^- \to e^+e^- \phi \to e^+e^- \eta' \gamma$ and $e^+e^- \to e^+e^- J/\psi \to e^+e^- \eta' \gamma$.
As in Ref.~\cite{old_etastudy}, the impact of these processes on the
  results is found to be negligible.
\section{\boldmath Detection efficiency}
\label{DETECT_EFFICIENCY_cross_section_and_form_factor}
\begin{table*}
\caption{The weighted averages $\overline{Q_1^2}$ and $\overline{Q_2^2}$ for the ($Q_1^2, Q_2^2$) region, the boundaries of the ($Q_1^2, Q_2^2$) region, the detection efficiency ($\varepsilon_{\rm true}$), the radiative correction
factor ($R$), the number of selected signal events ($N_{\rm events}$), the
cross section ($d^2\sigma(\overline{Q_{1}^2},\overline{Q_{2}^2})/(dQ_1^2dQ_2^2)$) with its statistical uncertainty,
and the  $\gamma^{\star}\gamma^{\star} \to \eta'$ transition form factor
($F(\overline{Q_1^2}$, $\overline{Q_2^2})$) with
the statistical, systematic, and model uncertainties (see text). All presented upper limits correspond to 90\% C.L..
\label{cross_table0}}
\begin{ruledtabular}
\begin{tabular}{cccccccc}
$\overline{Q_1^2}$, $\overline{Q_2^2}$ (GeV$^2$) & 
($Q_1^2, Q_2^2$) region (GeV$^2$)  & $\varepsilon_{\rm true}$ & $R$ & 
$N_{\rm events}$ & $d^2\sigma/(dQ_1^2dQ_2^2)$ &
$F(\overline{Q_1^2}$, $\overline{Q_2^2}$)  \\
 & & & & & $\times 10^4$ (fb/GeV$^4$) &$\times 10^3$ (GeV$^{-1}$) \\
\hline    
6.48, 6.48 & $2 < Q_1^2,Q_2^2 < 10 $ & 0.019 & 1.03  &14.7$^{+4.3}_{-3.6}$ & 1471.8$^{+430.1}_{-362.9}$ &   14.32$^{+1.95}_{-1.89} \pm $ 0.83 $\pm$ 0.14\\
16.85, 16.85 & $10 < Q_1^2,Q_2^2 < 30 $ & 0.282 & 1.10 & 4.2$^{+3.1}_{-2.7}$ & 4.2$^{+3.1}_{-2.7}$ &   5.35$^{+1.71}_{-2.15} \pm $ 0.31 $\pm$ 0.42\\
             &                          &       &       & $< 9.8$           & $<10.0$              &   $< 14.53$                                           \\
14.83, 4.27 & $10 < Q_1^2 < 30; 2 < Q_2^2 < 10$ & 0.145 & 1.07 & 15.8$^{+4.8}_{-4.0}$ & 39.7$^{+12.0}_{-10.2}$ &   8.24$^{+1.16}_{-1.13} \pm $ 0.48 $\pm$ 0.65\\
38.11, 14.95 & $30 < Q_1^2 < 60; 2 < Q_2^2 < 30$ & 0.226 & 1.11 & 10.0$^{+3.9}_{-3.2}$ & 3.0$^{+1.2}_{-1.0}$ &   6.07$^{+1.09}_{-1.07} \pm $ 0.35 $\pm$ 1.21\\
45.63, 45.63 & $30 < Q_1^2,Q_2^2 < 60$  & 0.293 & 1.22 & 1.6$^{+1.8}_{-1.1}$ & 0.6$^{+0.7}_{-0.6}$ &   8.71$^{+3.96}_{-8.71} \pm $ 0.50 $\pm$ 1.04\\
             &                         &        &       &     $<$ 5.0           &    $<$ 1.9                 &     $<$ 32.03                   \\
\end{tabular}
\end{ruledtabular}
\end{table*}
\begin{figure*}
\begin{center}
\includegraphics[width=.55\textwidth]{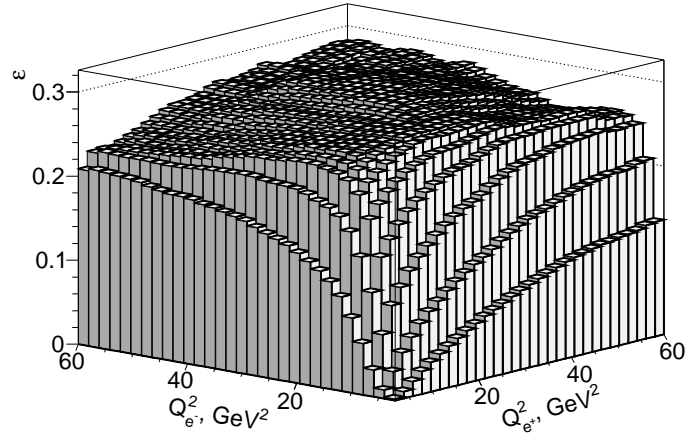} 
\caption{Detection efficiency as a function of the momentum transfers $Q_{e^-}^2$ and $Q_{e^+}^2$.
\label{efficiency}}
\end{center}
\end{figure*}
\begin{table}
\caption{The sources of the systematic uncertainties in the $e^+e^-\to e^+e^-\eta'$ cross section. 
\label{systemtable}}
\begin{ruledtabular}
\begin{tabular}{lc}
Source & Uncertainty (\%)\\
\hline
$\pi^{\pm}$ identification & 1.0 \\
$e^{\pm}$ identification & 1.0 \\
Other selection criteria & 11.0 \\
Track reconstruction & 0.9 \\
$\eta \to 2\gamma$ reconstruction & 2.0 \\
Trigger, filters   &  1.3 \\
Background subtraction & 3.7 \\
Radiative correction & 1.0 \\
Luminosity & 1.0 \\
\hline
Total & \systcr \\ 
\end{tabular}
\end{ruledtabular}
 \end{table}
 
The detection efficiency ($\varepsilon$) is
determined from MC simulation in the ($Q^2_{e^-}$, $Q^2_{e^+}$) plane as the ratio
of the selected over generated events and is shown in Fig.~\ref{efficiency}.
The detector acceptance limits the efficiency at small momenta and the minimum  measurable $Q^{2}$ is 2 GeV$^2$. The difference 
between the energies of the $e^+$ and $e^-$ beams at PEP-II leads to an
asymmetry in the dependence of the efficiency on $Q^2_{e^+}$ and $Q^2_{e^-}$.

Because of the symmetry of the form factor 
$F_{\eta'}(Q_1^2, Q_2^2) = F_{\eta'}(Q_2^2, Q_1^2)$,
 we use the notation
\begin{linenomath}
\begin{equation}\label{FsdsdsdF}
Q_{1}^2 = \mathrm{max}(Q_{e^{+}}^2,Q_{e^{-}}^2),\:
Q_{2}^2 = \mathrm{min}(Q_{e^{+}}^2,Q_{e^{-}}^2).
\end{equation}
\end{linenomath}
Since signal MC events are generated with a constant TFF, the average 
detection efficiency for the specific ($Q_1^2, Q_2^2$) region is calculated as
the ratio of the following integrals:
\begin{linenomath}
\begin{equation}\label{lefftrue}
\varepsilon_{true} = \frac {\int \varepsilon(Q_{1}^2,Q_{2}^2) F^2_{\eta'}(Q_{1}^2,Q_{2}^2) dQ_{1}^2dQ_{2}^2} {\int F^2_{\eta'}(Q_{1}^2,Q_{2}^2) dQ_{1}^2dQ_{2}^2},
\end{equation}
\end{linenomath}
where the form factor is described by Eq.~(\ref{adsadasdasd}).
The obtained values of the detection efficiency for the five ($Q_1^2, Q_2^2$)
regions are listed in Table~\ref{cross_table0}.

The  systematic  uncertainties related to the detection efficiency are listed in Table~\ref{systemtable}.
The uncertainties related to track reconstruction, $\eta \to 2\gamma$ reconstruction, trigger and filters, and the pion PID were
studied in our previous single-tag analysis~\cite{old_etastudy}.
To estimate the efficiency uncertainty related to 
other selection criteria, we apply a less strict 
condition on a criterion, perform the procedure of 
background subtraction described in the previous section, 
and calculate the ratio of the number of selected events 
in data and simulation.
We consider the less strict requirements $P^{\star}_{e^+e^-\eta'} < 1$~GeV/$c$, 
$10.20 < E^{\star}_{e^+e^-\eta'} < 10.75$ GeV, $ 0.48 < M_{\gamma\gamma} < 0.60$ GeV/$c^2$, and remove the requirements on $E_{e^+}$ and $E_{e^-}$ entirely. The quadratic sum of the deviations from the nominal value of the ratio (11\%)  is used as the total systematic uncertainty of the detection efficiency. 
\section{\boldmath Cross section and form factor
\label{ldiscussion}}
The differential Born cross section for the process 
$e^+e^-\to e^+e^- \eta^{\prime}$ is calculated as
\begin{linenomath}
\begin{equation}\label{l2621}
\frac{d^2\sigma}{dQ_{1}^2dQ_{2}^2} = \frac{1}{\varepsilon_{\rm true} R \mathcal{L} \mathcal{B}}
\frac{d^2N}{dQ_{1}^2dQ_{2}^2},
\end{equation}
\end{linenomath}
where $d^2N/(dQ_{1}^2dQ_{2}^2)$ is the number of signal events in the
($Q_1^2, Q_2^2$) region divided by the area of this region, $\mathcal{L}$ is the integrated luminosity, and $R$ is a radiative correction factor accounting for distortion of the $Q^2_{1,2}$ spectrum due to the emission of photons from the initial state and for vacuum polarization effects.  
The factor $\mathcal{B}$ is the product of the branching fractions
$\mathcal{B}(\eta^{\prime}\to \pi^+ \pi^- \eta)\mathcal{B}(\eta \to \gamma\gamma)=0.169\pm0.003$~\cite{PDG2014}.
The radiative correction factor $R$ is determined using simulation at the
generator level, i.e., without detector simulation. The $Q^2_{1,2}$  spectrum is 
generated using only the pure Born amplitude for the 
$e^+e^-\to e^+e^- \eta'$ process, and then using a model with radiative 
corrections included. The factor $R$ is evaluated as the ratio of the second
spectrum to the first. 
The values of the cross section for the five ($Q_1^2, Q_2^2$) regions are
listed in Table~\ref{cross_table0}.
The cross section in the entire range of momentum transfer
$2 < Q_1^2, Q_2^2 < 60$ GeV$^2$ is
\begin{linenomath}
\begin{equation}\label{crossss}
\sigma = 11.4^{+2.8}_{-2.4}~\rm fb, 
\end{equation}
\end{linenomath}
where the uncertainty is statistical. The systematic uncertainty includes the
uncertainty in the number of signal events associated with background subtraction
(Sec.~\ref{eventselection}), the uncertainty in the detection efficiency
(Sec.~\ref{DETECT_EFFICIENCY_cross_section_and_form_factor}), the uncertainty
in the calculation of the radiative correction (1\%)~\cite{abrakadabra},
and the uncertainty in the integrated luminosity (1\%)~\cite{lumi}.
All sources of systematic uncertainty in the cross section
are summarized in Table~\ref{systemtable}. The total systematic uncertainty
(\systcr) ~is the sum in quadrature of all the systematic contributions. The model uncertainty will be discussed below.

To extract the TFF we compare the value of the measured cross 
section  from Eq.~(\ref{l2621}) with the calculated one. The latter is evaluated using
$F_{\eta'}^{2}(Q_1^2,Q_2^2)$ obtained from Eq.~(\ref{adsadasdasd}).
Therefore, the measured form factor is determined as
\begin{linenomath}
\begin{equation}\label{l23} 
F^2(\overline{Q_1^2},\overline{Q_2^2}) = \frac{(d^2\sigma/(dQ_1^2dQ_2^2))_{data}}{(d^2\sigma/(dQ_1^2dQ_2^2))_{MC}}F_{\eta'}^{2}(\overline{Q_1^2},\overline{Q_2^2}), 
\end{equation}
\end{linenomath}
where $F_{\eta'}^2(\overline{Q_1^2},\overline{Q_2^2})$ and $(d^2\sigma/(dQ_1^2dQ_2^2))_{MC}$ correspond to Eq.~(\ref{adsadasdasd}).

The average momentum transfer squared for each ($Q_1^2, Q_2^2$) region is 
calculated using the data spectrum normalized to the detection efficiency:
\begin{linenomath}
\begin{equation}\label{eq_Q_averaging}
\overline{Q_{1,2}^{2}} = \frac{\sum_{i} Q^2_{1,2}(i)/\varepsilon(Q_1^2, Q_2^2)}{\sum_{i} 1/\varepsilon(Q_1^2, Q_2^2)}.
\end{equation}
\end{linenomath}
For regions 1, 2, and 5, the $\overline{Q_{1}^{2}}$ and $\overline{Q_{2}^{2}}$ are additionally averaged. 

\begin{figure}
\includegraphics[width=0.49\textwidth]{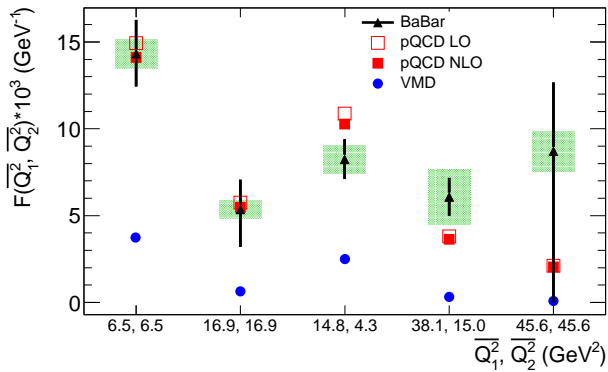} 
\caption{Comparison of the measured 
$\gamma^{\star}\gamma^{\star} \to \eta'$ transition form factor (triangles, with
error bars representing the statistical uncertainties) with the LO (open squares)
and NLO (filled squares) pQCD predictions and the VDM predictions (circles).
\label{formfactor_compare}}
\end{figure}

The model uncertainty arises from the model dependence of  
$(d^2\sigma/(dQ_1^2dQ_2^2))_{MC}$ and $\varepsilon_{true}$. 
Repeating the calculation of Eqs.~(\ref{lefftrue}),~(\ref{l2621}), and~(\ref{l23}) with a constant 
TFF, we estimate the model uncertainty. In the case of the cross section it is about 
60\% because of the strong dependence of $\varepsilon_{true}$ on the input model 
for TFF at small values of $Q_1^2$ and $Q_2^2$. However, the transition form
factor is much less sensitive to the model.

The obtained values of the transition form factor are listed in 
Table~\ref{cross_table0} and are represented in 
Fig.~\ref{formfactor_compare} 
by the triangles. The error bars attached to the triangles indicate the statistical uncertainties.
  The quadratic sum of the systematic and model uncertainties is shown by the shaded rectangles. The open and filled squares in Fig.~\ref{formfactor_compare}
correspond to the LO and NLO pQCD predictions [Eq.~(\ref{adsadasdasd})], 
respectively. The NLO correction is relatively small.
The measured TFF is, in general, consistent with the QCD prediction. The 
circles in Fig.~\ref{formfactor_compare} represent the predictions of the
VDM model [Eq.~(\ref{lfff8})], which exhibits a clear disagreement with the data.
\section{\boldmath Summary}
\label{lSummary}
We have studied for the first time the process $e^+e^-\to e^+e^- \eta'$ 
in the double-tag mode and have measured the 
$\gamma^{\star}\gamma^{\star} \to \eta'$ transition form factor
in the momentum-transfer range $2 < Q_1^2, Q_2^2 < 60$ GeV$^2$.
The measured values of the form factor are in agreement with the pQCD prediction
and contradict the prediction of the VDM model. 
\section{ACKNOWLEDGMENTS}

We are grateful for the extraordinary contributions of our \pep2\ colleagues in
achieving the excellent luminosity and machine conditions
that have made this work possible. The success of this project also relies critically on the 
expertise and dedication of the computing organizations that 
support \babar.
The collaborating institutions wish to thank 
SLAC for its support and the kind hospitality extended to them. 
This work is supported by the
US Department of Energy
and National Science Foundation, the
Natural Sciences and Engineering Research Council (Canada),
the Commissariat \`a l'Energie Atomique and
Institut National de Physique Nucl\'eaire et de Physique des Particules
(France), the
Bundesministerium f\"ur Bildung und Forschung and
Deutsche Forschungsgemeinschaft
(Germany), the
Istituto Nazionale di Fisica Nucleare (Italy),
the Foundation for Fundamental Research on Matter (The Netherlands),
the Research Council of Norway, the
Ministry of Education and Science of the Russian Federation, 
Ministerio de Econom\'{\i}a y Competitividad (Spain), the
Science and Technology Facilities Council (United Kingdom),
and the Binational Science Foundation (U.S.-Israel).
ndividuals  have  received  support
from the Russian Foundation for Basic Research (grant
No.  18-32-01020), the Marie-Curie IEF program (Euro-
pean Union) and the A. P. Sloan Foundation (USA).


\begin{thebibliography}{0}
 \bibitem{bPLUTO}
C. Berger {\it et al.} (PLUTO Collaboration), Phys. Lett. B {\bf 142}, 125 (1984). 

 \bibitem{bTPC}
H. Aihara {\it et al.} (TPC/Two Gamma Collaboration), Phys. Rev. D {\bf 38}, 1 (1988); Phys. Rev. Lett. {\bf 64}, 172 (1990).

 \bibitem{bCELLO}
H.-J. Behrend {\it et al.} (CELLO Collaboration), Z. Phys. C {\bf 49} (1991) 401.

 \bibitem{bCLEO}
 J. Gronberg  {\it et al.} (CLEO Collaboration), Phys. Rev. D {\bf 57}, 33 (1998).
 
\bibitem{old_etastudy}
P.~del Amo Sanchez {\it et al.} (\babar ~Collaboration),
Phys.\ Rev.\ D {\bf 84}, 052001 (2011).
 

\bibitem{QCD_inspired_model}
G. Kopp, T. F. Walsh, and P. M. Zerwas, Nucl. Phys. B {\bf 70}, 461 (1974).
\bibitem{bVMD1}
S. Berman and D. Geffen, Nuovo Cim. {\bf 18}, 1192 (1960). 
\bibitem{bKroll}
P. Kroll, Nucl. Phys. B (Proc. Suppl.) {\bf 219-220}, 2 (2011).
\bibitem{bsumrules}
S. Agaev  {\it et al.}, Phys. Rev. D {\bf 90}, 074019 (2014).
\bibitem{bVMD2}
B.-l. Young, Phys. Rev. {\bf 161}, 1620 (1967).

\bibitem{bVMD}
L. G. Landsberg, Phys. Rep. {\bf 128}, 301 (1985).

\bibitem{bVMD3}
A. Dorokhov, M. Ivanov, and S. Kovalenko, Phys.Lett. B {\bf 677}, 145 (2009).

\bibitem{et_etap_mixing} 
T. Feldmann, P. Kroll and B. Stech, Phys. Rev. D {\bf 58}, 114006 (1998).

\bibitem{et_etap_mixing1}
Fu-Guang Cao, Phys. Rev. D {\bf 85}, 057501 (2012).

\bibitem{BRAATEN}
E. Braaten, Phys. Rev. D {\bf 28}, 524 (1983). 

\bibitem{Brodsky_Lepage}
S. J. Brodsky and G. P. Lepage, Phys. Rev. D {\bf 24}, 7 (1981);
G. P. Lepage and S. J. Brodsky, Phys. Rev. D {\bf 22}, 2157 (1980).

\bibitem{Chernyak} V. L. Chernyak and A. R. Zhitnitsky, Nucl. Phys. B {\bf 201}, 492 (1982); Phys. Rep. {\bf 112} 173 (1984); Nucl. Phys. B {\bf 246}, 52 (1984).

\bibitem{TFF_in_Q2_limit} 
T. Feldmann and P. Kroll, Phys. Rev. D {\bf 58}, 057501 (1998).
\bibitem{Brodsky2011}
A. J. Brodsky, F. Cao and G. Teramond, Phys. Rev. D {\bf 84}, 033001 (2011).
\bibitem{PDG2014}
C. Patrignani {\it et al.} (Particle Data Group), Chin. Phys. C {\bf40}, 100001 (2016).
\bibitem{lumi}J. P. Lees {\it et al.} (\babar ~Collaboration), 
Nucl. Instrum. Methods Phys. Res., Sect. A {\bf726}, 203 (2013).
\bibitem{Detector}
B.~Aubert {\it et al.} ({\babar} Collaboration), Nucl. Instrum. and Meth. A {\bf 479}, 1 (2002).
\bibitem{Detector1}
B.~Aubert {\it et al.} ({\babar} Collaboration), Nucl. Instrum. and Meth. A {\bf 729}, 615 (2013).
  
\bibitem{rGGResRc}
V.~P.~Druzhinin, L.~V.~Kardapoltsev and V.~A.~Tayursky,
Comput.\ Phys.\ Commun.\  {\bf 185}, 236 (2014).
\bibitem{abrakadabra}
S. Ong and P. Kessler, Phys. Rev. D  {\bf38}, 2280 (1988).
\bibitem{Feldman}
G. J. Feldman and R. D. Cousins, Phys. Rev. D {\bf57}, 3873 (1998).
 
 \end{thebibliography}
\end{document}